\documentclass{waica}
\usepackage[T1]{fontenc}
\usepackage[utf8]{inputenc}
\usepackage{graphicx}
\usepackage{booktabs}
\usepackage{amsmath,amsfonts,amssymb,bm}
\usepackage{xcolor}
\usepackage{enumitem}
\usepackage{adjustbox}
\usepackage{array}
\usepackage{makecell}
\usepackage{caption}
\usepackage{subcaption}
\usepackage{float}
\usepackage[section]{placeins}
\usepackage{tikz}
\usetikzlibrary{arrows.meta,positioning,fit,calc}
\usepackage{lmodern}
\usepackage[final,activate={true,nocompatibility},factor=1100,stretch=10,shrink=10]{microtype}
\usepackage{multirow}
\usepackage{cite}
\usepackage{hyperref}
\usepackage{color}

\usepackage{cleveref}

\let\citep\cite
\let\citet\cite

\newcommand{\ourmodel}{\mbox{ParaASR}}
\newcommand{\mtpmodel}{MTP-5}

\setlist[itemize,1]{leftmargin=18pt}
\setlist[enumerate,1]{leftmargin=18pt}
\begin{document}
\title{\ourmodel: Multi-Token Prediction for Fast and Long-Context LLM-Based Speech Recognition}
%
%
\author{Qingjian~Lin\inst{1} \and
Yuxin~Li\inst{1,2} \and
Haoyang~Zhang\inst{1,3} \and
Jun~Chen\inst{1} \and
Yechang~Huang\inst{1} \and
Feng~Tian\inst{1} \and
Xie~Li\inst{1} \and
Xiangyu~Tony~Zhang\inst{1,4} \and
Daijiao~Liu\inst{1,4} \and
Yuxin~Zhang\inst{1,5} \and
Jinglan~Gong\inst{1,6} \and
Bo~Zhao\inst{1} \and
Fei~Tian\inst{1}\thanks{Corresponding author.} \and
Xuerui~Yang\inst{1} \and
Gang~Yu\inst{1} \and
Xiangyu~Zhang\inst{1} \and
Daxin~Jiang\inst{1}}
\authorrunning{Q. Lin et al.}
\institute{%
$^{1}$StepFun \quad
$^{2}$NTU \quad
$^{3}$PKU \quad
$^{4}$UNSW \quad
$^{5}$SJTU \quad
$^{6}$USTC \\[2pt]
\email{linqingjian@stepfun.com, tianfei@stepfun.com}}
\maketitle

\begin{abstract}
Audio-encoder--LLM-decoder architectures have emerged as the dominant paradigm for modern automatic speech recognition (ASR), substantially improving transcription quality through large-scale language modeling. However, the computational cost of autoregressive decoding scales with decoder size, creating a fundamental trade-off between recognition quality and serving latency. This raises a central question: \emph{can LLM-based ASR attain state-of-the-art accuracy without sacrificing efficiency?} We argue that this trade-off is not inherent to the problem itself. Unlike open-ended text generation, ASR outputs are strongly anchored to the input speech signal, providing a natural inductive bias toward high-parallelism decoding. Building on this observation, we introduce \ourmodel, an ASR system that leverages Multi-Token Prediction (MTP) to let a 4B LLM decoder emit multiple tokens per forward step. Starting from a publicly available audio-language foundation, the model first establishes a robust autoregressive recognizer and then aligns five future-token branches through a staged optimization recipe. At inference time, the system proposes a six-token continuation per step and admits only the verified prefix into the transcript, thus preserving the safety of standard autoregressive decoding. The average accepted length reaches 5.0 out of 6 proposed tokens, confirming that the deterministic structure of speech makes ASR an especially natural setting for multi-token decoding. \ourmodel\ further retains a native 32K-context window and transcribes up to 30 minutes of audio in a single pass. Across diverse benchmarks, the model attains average error rates of 2.97\%, 3.68\%, and 3.70\% on Chinese, English, and long-form evaluations, respectively, while reaching a real-time factor (RTF) as low as 0.0053. Together, these results show that decoder scaling, low-latency inference, and long-context transcription need not be competing goals when future-token proposals are anchored by the acoustic signal and guarded by autoregressive verification.

\end{abstract}

\section{Introduction}

\begin{figure}[!t]
    \centering
    \includegraphics[width=\linewidth]{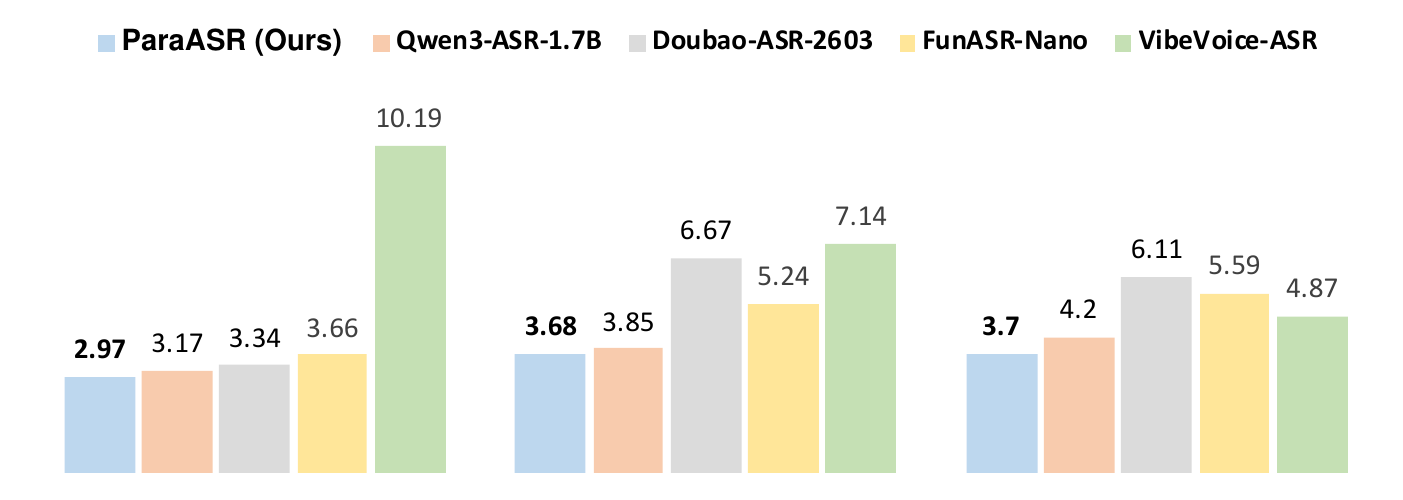}
    \caption{Performance comparison across Chinese, English, and long-form ASR benchmarks. Chinese results are reported in CER, while English and long-form results are reported in WER. Lower is better.}
    \label{fig:performance_overview}
\end{figure}

Automatic speech recognition (ASR) is gradually shifting from acoustic sequence transduction~\cite{graves2012connectionist, chan2015listen, graves2012sequence, zhang2025mamba} toward audio-grounded language generation~\cite{malik2021automatic}. Under this formulation, an audio encoder extracts speech evidence and a language-model decoder writes the transcript conditioned on that evidence. Such a shift is appealing because many recognition errors are not purely acoustic in nature: resolving homophones, handling code-switching~\cite{liu2025code}, restoring punctuation, normalizing named entities and domain terminology, and maintaining long-range consistency all benefit from stronger language modeling. This paradigm has been explored by large-scale weakly supervised recognizers such as Whisper~\cite{radford2023robust}, by audio-language foundations like StepAudio~\cite{wu2025stepaudio2technicalreport,lin2026stepaudio,tian2025step,zhang2026step} and Qwen3-Omni~\cite{xu2025qwen3omnitechnicalreport}, by real-time spoken language models with integrated reasoning~\cite{wu2025mind}, by full-duplex interactive systems and dedicated conversational benchmarks~\cite{zhang2026duplexsla,deng2025multi}, and by a growing family of production-oriented LLM-based ASR systems~\cite{shi2026qwen3asrtechnicalreport,an2025funasrtechnicalreport,peng2026vibevoiceasrtechnicalreport,bai2024seed}.

This formulation, however, introduces a non-trivial deployment bottleneck. A larger decoder provides the linguistic depth needed for high-quality transcription, but its computational cost is incurred for \emph{every} output token. The penalty becomes particularly severe in long-form ASR, where extended recordings simultaneously demand high throughput and long-range consistency~\cite{bain2023whisperx, koluguri2024investigating}. \ourmodel\ resolves this tension by moving beyond strictly sequential generation: it exploits the deterministic nature of the ASR task to enable parallel token emission while preserving the safety guarantees of autoregressive decoding.

Unlike open-ended generation, speech transcription is tightly anchored to acoustic evidence. Given the audio signal and the preceding transcript, the local future is largely a matter of faithful continuation rather than open-ended choice~\cite{bai2024seed}. Multi-token prediction and speculative decoding methods exploit a similar principle in LLM inference by proposing multiple tokens and verifying them against a target distribution~\cite{leviathan2023fast,chen2023accelerating,cai2024medusa,gloeckle2024betterfaster,huang2026step35flashopen}. While these methods are often bottlenecked by semantic branching in free-form text generation, the speech signal in ASR makes lookahead proposals remarkably predictable, effectively converting the deterministic structure of speech into a high-parallelism decoding advantage. Autoregressive verification then ensures that the final transcript remains as reliable as the one produced by standard decoding.

Building on this insight, \ourmodel\ integrates MTP-based parallel decoding into an LLM-based ASR system. The model couples a frozen audio encoder, a linear adapter, and a dense Transformer decoder, augmented with MTP branches that propose future tokens in parallel with the main branch. This design allows the system to emit multiple verified tokens per forward step, effectively pushing the quality--efficiency frontier of modern ASR by delivering the accuracy of a large-scale decoder with the serving efficiency required for real-time applications.

\begin{figure}[!t]
\centering
\includegraphics[width=\linewidth]{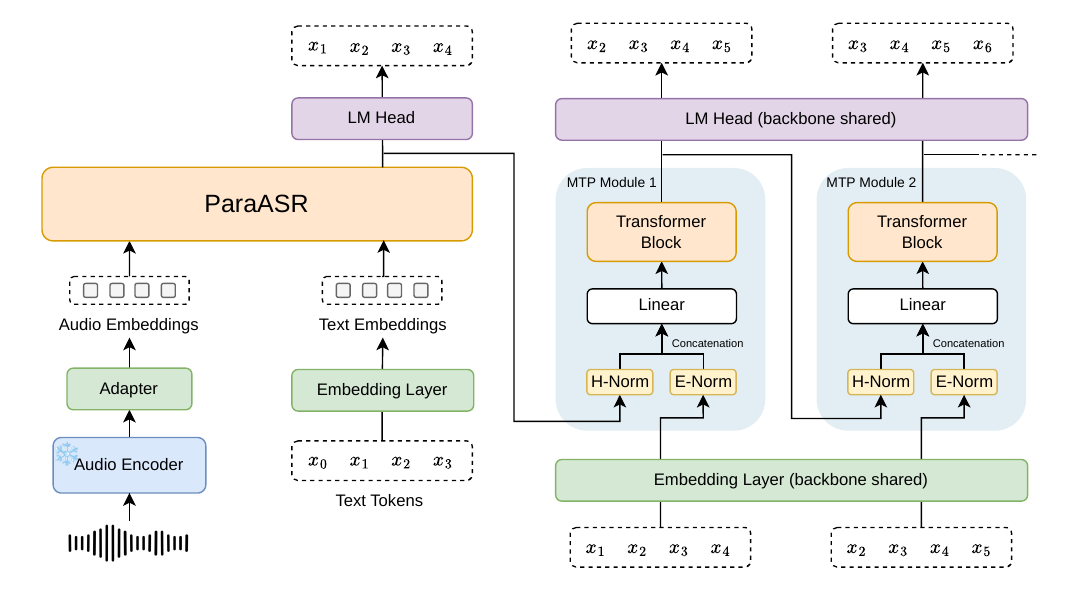}
\caption{Architecture of \ourmodel. A frozen audio encoder produces 80\,ms acoustic embeddings, a linear adapter maps them to the decoder hidden space, and multiple MTP blocks propose future transcript tokens in parallel with the main next-token prediction.}
\label{fig:architecture}
\end{figure}

The training recipe is deliberately structured to preserve recognition stability while enabling acceleration. Audio-language pretraining first establishes the multimodal foundation, after which ASR supervised fine-tuning builds a reliable autoregressive recognizer. Only then is a staged MTP optimization process applied to align the future-token branches with the converged recognizer. This ordering ensures that MTP acts as a calibrated acceleration module rather than a competing objective. The resulting model retains a native 32K-context window and supports up to 30 minutes of audio in a single decoding session, making it well suited to complex long-form scenarios such as meetings and broadcasts.

Extensive experiments show that \ourmodel\ noticeably advances the quality--efficiency frontier of LLM-based ASR. As summarized in Figure~\ref{fig:performance_overview}, the model attains strong recognition performance across Chinese, English, and long-form benchmarks relative to competitive ASR baselines. At the same time, it reaches an exceptionally low RTF of 0.0053 under single-GPU serving. On the WenetSpeech meeting subset~\cite{zhang2022wenetspeech}, the average accepted length reaches 5.0 out of 6 proposed tokens, while matched ablations confirm that MTP-based acceleration leaves recognition accuracy essentially unchanged. Together, these findings support the central premise of this paper: ASR is an especially natural application scenario for multi-token prediction, since speech makes future tokens predictable enough for aggressive acceleration, and autoregressive verification keeps the final transcript reliable.

The contributions of this paper are summarized as follows:
\begin{itemize}
    \item We identify the determinism of speech transcription as a key opportunity for accelerating LLM-based ASR, and formulate MTP as a natural decoding paradigm in this setting.
    \item We present \ourmodel, a 4B-decoder ASR system with five MTP branches and native 32K-context support tailored for long-audio transcription.
    \item Through comprehensive evaluation, we show that decoder scaling, low-latency inference, and long-context transcription need not be opposing goals in ASR.
\end{itemize}

\section{Model Overview}

\ourmodel\ is intentionally conservative in its backbone but aggressive in its decoding path. The architecture follows an encoder--adapter--decoder pattern, augmented with an \mtpmodel\ head that proposes verifiable future transcript tokens, as illustrated in Figure~\ref{fig:architecture}. The design is organized around three concurrent goals: strong recognition accuracy from a large decoder, efficient serving through multi-token proposals, and native long-form transcription enabled by a 32K context window.

\subsection{Backbone Architecture}

The backbone follows the standard encoder--adapter--decoder pattern for audio-language modeling. The audio encoder is a 0.6B Transformer initialized from a publicly available omni-modal foundation~\cite{xu2025qwen3omnitechnicalreport}. It remains frozen throughout training and applies $8\times$ temporal downsampling to produce one acoustic embedding every 80\,ms; the frame rate and token granularity of speech representations are known to affect recognition quality in speech language models~\cite{zhang2025impact,zhang2026trap}. These embeddings are projected by a linear adapter into the hidden space of the language decoder. The decoder itself is a 4B dense Transformer initialized from a pre-trained text LLM and supports a native 32K context window. Combined with the 80\,ms acoustic embedding rate, this large context budget allows \ourmodel\ to transcribe up to 30 minutes of audio in a single decoding session, removing the need for complex chunk-and-stitch pipelines.

\subsection{MTP as Verifiable Lookahead}

At decoding position $t$, the main branch predicts the next transcript token $x_{t+1}$. The $h$-th MTP branch predicts $x_{t+1+h}$ for $h\in\{1,\ldots,5\}$, so a single forward step yields a six-token proposal. At inference time, the proposal is admitted only as a verified prefix: once a future token disagrees with the normal decoding path, all subsequent proposed tokens are rejected and decoding continues autoregressively from the accepted prefix. This verification step ensures that MTP acts strictly as an acceleration primitive: it improves serving efficiency by increasing the tokens emitted per step, while preserving the safety rule that ultimately determines the transcript.

Each MTP block takes as input the hidden state from the previous branch together with a shifted token embedding. The two are normalized, concatenated, projected back to the decoder hidden size, and processed by a decoder-style Transformer block, as shown in Figure~\ref{fig:architecture}. All branches share the same embedding layer and vocabulary output head as the main decoder. This structural alignment keeps the lookahead proposals linguistically consistent with the main autoregressive path.

\section{Data Preparation and Training Recipe}

The training recipe follows a staged post-training path~\cite{liu2026boosting}: from foundation learning, to ASR specialization, and finally to decoding acceleration. Audio-language pretraining provides the speech-to-language foundation. ASR supervised fine-tuning (SFT) is then carried out without Multi-Token Prediction (MTP), using a mixture of short-form and long-form transcription data to build a reliable autoregressive recognizer. Only after this recognizer has fully converged do we attach the MTP branches for future-token prediction. This ordering keeps MTP as an acceleration module rather than a new recognition objective: it preserves the stability and accuracy of the main ASR decoder while enabling verified multi-token proposals at inference time.

\subsection{Inherited Audio-Language Pretraining}

We initialize from a publicly available audio-language pretraining recipe~\cite{wu2025stepaudio2technicalreport}, which establishes a robust foundation over 1.356T text and audio tokens. This process equips the model with strong audio understanding and generation capabilities while preserving the linguistic competence of the underlying LLM. The pretraining proceeds in four progressive phases:
\begin{itemize}
\item \textbf{Speech--text alignment.} 100B ASR tokens are used to align the audio adaptor with the LLM text embedding space. During this 12K-step phase, both the audio encoder and the LLM remain frozen to ensure a stable feature mapping.
\item \textbf{Audio-token extension.} The tokenizer is extended with 6.6K discrete audio tokens. The model is then trained on 128B text and 128B audio tokens (including TTS and speech-to-speech data) to support unified multimodal modeling while preserving textual capabilities.
\item \textbf{Unified multimodal pretraining.} This main stage scales the model with 800B tokens. The mixture includes 400B text tokens alongside diverse audio tasks such as ASR, TTS, speech translation, and interleaved text--speech continuation, facilitating deep cross-modal reasoning.
\item \textbf{Cooldown and capability expansion.} 200B high-quality tokens are used to introduce paralinguistic understanding and further refine multilingual ASR. A conversational speech synthesis pipeline with 50K unique speakers is leveraged to ensure vocal diversity and emotional expressiveness.
\end{itemize}

Through this multi-stage pretraining, the model acquires a comprehensive multimodal representation, allowing the subsequent ASR specialization to benefit from both semantic knowledge and acoustic paralinguistic cues.

\subsection{ASR Supervised Fine-Tuning}

ASR SFT adapts the pretrained foundation into a dedicated transcription model. By combining short-form supervision for utterance-level precision with long-form supervision for contextual consistency, the recipe enables a single recognizer to handle diverse recording lengths and scenarios.

\textbf{Short-form supervised data.} The short-form SFT set comprises approximately 100K hours of audio, integrating major public corpora with extensive proprietary datasets. The mixture spans a wide spectrum of linguistic and acoustic variations, including Mandarin, English, and frequent code-switching utterances. We place particular emphasis on regional diversity by incorporating major Chinese dialects and regionally accented Mandarin. To handle real-world complexity, the data also covers vertical domains rich in professional terminology, as well as challenging acoustic environments such as far-field recordings and high-noise scenarios. To guarantee label quality, the proprietary data is manually verified to ensure high-fidelity alignment between audio and transcripts. Each sample in this set is at most 30 seconds long.

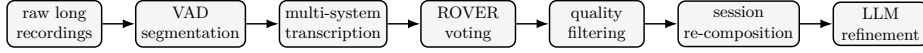
\begin{figure}[!tbp]
\centering
\resizebox{\textwidth}{!}{%
\begin{tikzpicture}[
    node distance=0.52cm and 0.62cm,
    box/.style={draw, rounded corners, thick, align=center, minimum height=0.7cm, minimum width=1.8cm, fill=gray!8},
    arrow/.style={-Latex, thick}
]
\node[box] (crawl) {raw long\\recordings};
\node[box, right=of crawl] (vad) {VAD\\segmentation};
\node[box, right=of vad] (asr) {multi-system\\transcription};
\node[box, right=of asr] (rover) {ROVER\\voting};
\node[box, right=of rover] (filter) {quality\\filtering};
\node[box, right=of filter] (concat) {session\\re-composition};
\node[box, right=of concat] (repair) {LLM\\refinement};

\draw[arrow] (crawl) -- (vad);
\draw[arrow] (vad) -- (asr);
\draw[arrow] (asr) -- (rover);
\draw[arrow] (rover) -- (filter);
\draw[arrow] (filter) -- (concat);
\draw[arrow] (concat) -- (repair);
\end{tikzpicture}%
}
\caption{Long-form ASR data construction pipeline. The process moves from individual clip transcription to global session-level refinement, ensuring both accuracy and consistency.}
\label{fig:longdata}
\end{figure}

\textbf{Long-form pseudo-labeled data.} While short-form data ensures utterance-level precision, long-duration recordings are essential for teaching the model to maintain contextual consistency. To support this capability, we curate a 50K-hour long-form dataset using a multi-system verification pipeline designed to provide reliable session-level supervision. Raw recordings are first segmented by Voice Activity Detection (VAD) into speech clips of at most 30 seconds. Each clip is independently transcribed by three ASR systems to obtain multiple candidate hypotheses. To focus the subsequent fusion on genuine recognition errors rather than surface variations, these hypotheses undergo surface-form normalization to unify formatting, casing, and punctuation. The normalized streams are then aligned and fused via Recognizer Output Voting Error Reduction (ROVER)~\cite{fiscus1997rover}, with voting performed at the character level for Chinese and at the word level for English, following the broader principle that adaptively combining multiple complementary estimators yields more reliable outputs than any single source~\cite{xuan2023new}. We accept tokens only when supported by at least two systems, and mark non-consensus positions as disagreements. The segment-level disagreement rate $\hat{e}$ then serves as a proxy for label reliability:
\[
\hat{e}=\frac{\#\mathrm{disagreed\ positions}}{\#\mathrm{text\ units}}.
\]
Clips with $\hat{e}>0.05$ are discarded to maintain high training signal fidelity. The remaining neighbor segments are concatenated into long-form training samples. Finally, an LLM-based refinement stage is applied to restore punctuation, perform Inverse Text Normalization (ITN), and—most importantly—ensure cross-segment consistency. By analyzing the entire session, the LLM harmonizes recurring terminology and entities that may otherwise vary across individual clips, producing a coherent and stable transcript for long-form supervision. The full pipeline, summarized in Figure~\ref{fig:longdata}, ensures that the model learns from high-quality, long-range acoustic--text pairs.

\textbf{Training details.}
ASR SFT follows a chat-style instruction format: the system prompt specifies the transcription task, the user input contains the audio features, and the assistant output emits the language tag followed by the transcript. To improve robustness under adverse acoustic conditions, non-speech or heavily noisy inputs are included with a non-speech tag and an empty transcript. Training examples are packed into a 32K-token sequence budget. SpecAugment-style time and frequency masking~\cite{park2019specaugment} is applied to the acoustic features, acting as a regularizer that promotes generalization to the acoustic mismatch encountered at deployment~\cite{zeng2023improving}. Throughout this stage, the audio encoder remains frozen, while the adapter and language decoder are optimized for 10K steps with a peak learning rate of $2\times10^{-5}$, a global batch size of 32, 100 warmup steps, and cosine decay to $1\times10^{-6}$.

\subsection{MTP Training}

MTP is introduced as a lookahead proposal module through a two-stage optimization recipe consisting of frozen-branch alignment and joint calibration. This staged design ensures that the auxiliary branches learn to forecast a stable target distribution without destabilizing the core recognition behavior. Both stages inherit the 32K sequence budget, the 32 global batch size, and the 10K-step training horizon from the ASR SFT stage.

\textbf{Frozen-branch alignment.} Five MTP blocks are appended to the converged ASR decoder. The Transformer layer inside each block is initialized from the last decoder layer to inherit a strong linguistic prior, while the branch-specific projections are randomly initialized. In this stage, only the MTP blocks are optimized with a peak learning rate of $2\times10^{-4}$; all other modules, including the shared token embeddings and the LM head, remain frozen. By decoupling the lookahead path from the recognition path, we prevent the newly initialized branches from injecting noise into the established autoregressive behavior.

\textbf{Joint calibration.} Once the branches have aligned with the ASR distribution, we unfreeze the adapter and the LLM decoder and perform joint optimization with a lower learning rate of $2\times10^{-5}$. This stage further reduces the residual mismatch between the backbone states and the lookahead branches, turning MTP into a calibrated proposal mechanism. Crucially, the optimization remains anchored to the transcription objective, ensuring that the multi-token proposals stay tightly coupled with the acoustic evidence.

During the above two-stage training, the main branch predicts the next token $x_{t+1}$ at position $t$, while the $h$-th MTP branch targets the future token $x_{t+1+h}$ for $h\in\{1,\ldots,H\}$. Masks are applied so that only transcript tokens contribute to the loss. The branch weights are exponentially decayed to reflect the serial dependency between MTP positions:
\[
w_h = \frac{\alpha^{h-1}}{\sum_{j=1}^{H}\alpha^{j-1}}, \quad H=5, \quad \alpha=0.9.
\]
At each position $t$, the final objective combines the standard next-token loss with the weighted MTP losses:
\[
\mathcal{L}_t = \mathrm{CE}(p_t, x_{t+1}) + \sum_{h=1}^{H} w_h \mathrm{CE}(p_{t,h}, x_{t+1+h}),
\]
where $p_t$ and $p_{t,h}$ denote the distributions from the main branch and the $h$-th auxiliary branch, respectively.

\section{Evaluation}

Our evaluation of \ourmodel\ focuses on three primary objectives: recognition accuracy across diverse languages and recording lengths, native long-form transcription capability, and inference efficiency under production-scale serving. We compare \ourmodel\ against several competitive baselines, namely VibeVoice-ASR~\cite{peng2026vibevoiceasrtechnicalreport}, FunASR-Nano~\cite{an2025funasrtechnicalreport}, Doubao-ASR-2603~\cite{bai2024seed}, and Qwen3-ASR-1.7B~\cite{shi2026qwen3asrtechnicalreport}.
To ensure a fair comparison, all models are deployed in a local environment using a single NVIDIA H800 GPU with single-concurrency serving, except for Doubao-ASR-2603, which is accessed through its official API. For baselines that do not natively support long-form audio, such as FunASR-Nano, we use VAD to segment the recordings into clips of at most 30 seconds.
Recognition benchmarks include AISHELL-1~\cite{bu2017aishell}, AISHELL-2 (iOS test)~\cite{du2018aishell2}, WenetSpeech~\cite{zhang2022wenetspeech}, FLEURS~\cite{conneau2022fleurs}, LibriSpeech~\cite{panayotov2015librispeech}, Common Voice~\cite{ardila2020commonvoice}, VoxPopuli cleaned AA~\cite{artificialanalysis2026voxpopulicleaned}, and Earnings22 cleaned AA~\cite{artificialanalysis2026earnings22cleaned}. Long-form evaluation includes the LibriSpeech long variants, Earnings22 cleaned AA, and a Wenet testnet long set that we construct from the WenetSpeech testnet by concatenating adjacent segments that share the same source session and have minimal inter-segment silence.

\subsection{Performance Benchmarks}

\textbf{Recognition performance.} Table~\ref{tab:asr_results} shows that \ourmodel\ establishes a new performance frontier for LLM-based ASR. On Chinese benchmarks, the model lowers the average CER to 2.97\%, with a notable reduction on AISHELL-1 to 0.71\% and a competitive 2.63\% on FLEURS zh. On English, \ourmodel\ reduces the average WER to 3.68\%, outperforming the competitive baselines and showing particular strength on LibriSpeech clean (1.38\%) and VoxPopuli cleaned AA (2.76\%). Long-form transcription is the regime in which decoder context and linguistic depth matter simultaneously, and \ourmodel\ reaches the best average long-form error rate of 3.70\%, a clear improvement over Qwen3-ASR-1.7B. The model is especially effective on the long LibriSpeech variants, where the native 32K context window enables consistent recognition without the boundary errors typical of segmentation-based pipelines. These results validate the central design claim: \ourmodel\ delivers superior recognition quality across diverse recording lengths, and its native long-context training further benefits extended speech without requiring a separate segmentation-time stitching strategy.

\begin{table}[H]
\centering
\small
\setlength{\tabcolsep}{3.6pt}
\caption{ASR results on Chinese, English, and long-form benchmarks (Error Rate, \%). Lower is better. The second-best results are underlined.}
\label{tab:asr_results}
\begin{adjustbox}{max width=\textwidth}
\begin{tabular}{llccccc}
\toprule
Category & Test set & VibeVoice-ASR & FunASR-Nano & Doubao-ASR-2603 & Qwen3-ASR-1.7B & \ourmodel \\
\midrule
\multirow{6}{*}{Chinese} & AISHELL-1 & 5.19 & 1.88 & 2.07 & \underline{1.49} & \textbf{0.71} \\
& AISHELL-2 ios & 5.10 & 2.61 & 2.70 & \underline{2.50} & \textbf{2.29} \\
& Wenet testnet & 14.79 & 5.30 & \textbf{4.03} & \underline{4.44} & 4.54 \\
& Wenet testmeeting & 17.09 & 5.31 & 5.09 & \textbf{4.66} & \underline{4.70} \\
& FLEURS zh & 8.77 & 3.19 & 2.83 & \underline{2.74} & \textbf{2.63} \\
\cmidrule(lr){2-7}
& Average & 10.19 & 3.66 & 3.34 & \underline{3.17} & \textbf{2.97} \\
\midrule
\multirow{6}{*}{English} & LibriSpeech clean & 2.30 & 1.80 & 2.94 & \underline{1.69} & \textbf{1.38} \\
& LibriSpeech other & 5.79 & 4.43 & 5.98 & \underline{3.57} & \textbf{3.16} \\
& Common Voice v11 en & 20.03 & 11.05 & 14.06 & \textbf{7.50} & \underline{7.57} \\
& FLEURS en & 5.20 & 4.96 & 6.74 & \textbf{3.23} & \underline{3.55} \\
& VoxPopuli cleaned AA & \textbf{2.38} & 3.97 & 3.61 & 3.28 & \underline{2.76} \\
\cmidrule(lr){2-7}
& Average & 7.14 & 5.24 & 6.67 & \underline{3.85} & \textbf{3.68} \\
\midrule
\multirow{5}{*}{Long-form} & LibriSpeech clean long & \underline{1.66} & 2.34 & 2.81 & 1.95 & \textbf{1.27} \\
& LibriSpeech other long & \underline{3.48} & 4.89 & 5.59 & 3.81 & \textbf{2.90} \\
& Wenet testnet long & 8.73 & 4.74 & \textbf{3.72} & \underline{4.15} & \underline{4.09} \\
& Earnings22 cleaned AA & \textbf{5.62} & 10.38 & 12.33 & 6.90 & \underline{6.52} \\
\cmidrule(lr){2-7}
& Average & 4.87 & 5.59 & 6.11 & \underline{4.20} & \textbf{3.70} \\
\bottomrule
\end{tabular}
\end{adjustbox}
\end{table}

\textbf{Decoding efficiency.} Table~\ref{tab:rtf} directly evaluates the deployment objective. We measure the Real-Time Factor (RTF) on 100 clips of 30 seconds each. \ourmodel\ reaches an exceptionally low RTF of 0.0053, faster than the Qwen3-ASR-1.7B baseline despite using a 4B decoder, and substantially faster than VibeVoice-ASR, FunASR-Nano, and Doubao-ASR-2603 under the same serving setup. This is the key systems consequence of MTP for ASR: decoder scale no longer translates linearly into token-by-token latency, since most steps emit several verified transcript tokens at once.

\begin{table}[H]
\centering
\small
\caption{RTF comparison.}
\label{tab:rtf}
\begin{tabular}{lccccc}
\toprule
Model & VibeVoice-ASR & FunASR-Nano & Doubao-ASR-2603 & Qwen3-ASR-1.7B & \ourmodel \\
\midrule
RTF & 0.1039 & 0.0591 & 0.0640 & 0.0094 & \textbf{0.0053} \\
\bottomrule
\end{tabular}
\end{table}

\subsection{Analysis and Ablation Study}

\textbf{MTP acceptance behavior.} We further evaluate the strict per-position acceptance rate of MTP on the WenetSpeech meeting set~\cite{zhang2022wenetspeech}. To determine the optimal lookahead horizon, we compare three configurations: MTP-3, MTP-5, and MTP-7. The results in Table~\ref{tab:acceptance} reveal two consistent trends. First, the acceptance rates of earlier positions are nearly invariant to the total number of branches, suggesting that each MTP head learns a stable, independent prediction task. Second, starting from the second position, the acceptance rate decays at a roughly constant factor of about $0.9$ per branch. While increasing the number of branches from three to five yields a substantial 39\% gain in average accepted length (from 3.6 to 5.0), the additional step to MTP-7 brings only a more modest 22\% improvement (reaching 6.1). This diminishing return is driven by the high failure rates of the sixth and seventh positions, which can approach 47\%. In production, such rejections frequently trigger KV cache rollbacks and interrupt the decoding stream, quickly offsetting the marginal benefit of a longer lookahead. We therefore choose MTP-5 as a deliberate efficiency--complexity trade-off: under this configuration, the model achieves an average of 5.0 verified tokens per forward step, providing the linguistic depth of a 4B decoder at a remarkably low real-time factor.

\begin{table}[H]
\centering
\small
\setlength{\tabcolsep}{8pt}
\caption{Strict per-position MTP acceptance rate and average accepted length.}
\label{tab:acceptance}
\begin{adjustbox}{max width=\textwidth}
\begin{tabular}{lccccccc|c}
\toprule
Config & 1st & 2nd & 3rd & 4th & 5th & 6th & 7th & Avg. Length \\
\midrule
MTP-3 & 0.96 & 0.88 & 0.80 & -- & -- & -- & -- & 3.6 / 4 \\
MTP-5 & 0.95 & 0.88 & 0.80 & 0.71 & 0.64 & -- & -- & 5.0 / 6 \\
MTP-7 & 0.96 & 0.88 & 0.80 & 0.72 & 0.65 & 0.59 & 0.53 & 6.1 / 8 \\
\bottomrule
\end{tabular}
\end{adjustbox}
\end{table}

\textbf{Ablation of MTP.} Table~\ref{tab:ablation} confirms that MTP behaves as a safe acceleration primitive. We provide a matched ablation comparing \ourmodel\ against the base ASR model after SFT but before MTP training. Across Chinese, English, and long-form benchmarks, the addition of MTP-5 leaves recognition accuracy essentially unchanged, with average fluctuations within 0.06 absolute points. This stability stems from our staged training recipe and from the autoregressive verification process, which guarantees that the final transcript is always determined by the verified path. Verification thus turns the future-token forecast into a safe serving primitive: correct proposals reduce latency, while incorrect proposals merely shorten the accepted prefix. This property makes MTP a low-risk way to deploy a larger decoder for ASR, simultaneously moving the system toward stronger language modeling and a lower real-time factor.

\begin{table}[H]
\centering
\small
\setlength{\tabcolsep}{4pt}
\caption{MTP ablation study.}
\label{tab:ablation}
\begin{adjustbox}{max width=\textwidth}
\begin{tabular}{llccc}
\toprule
Category & Test set & w/o MTP & MTP-5 & $\Delta$ \\
\midrule
\multirow{6}{*}{Chinese} & AISHELL-1 & 0.79 & 0.71 & -0.08 \\
& AISHELL-2 ios & 2.30 & 2.29 & -0.01 \\
& Wenet testnet & 4.57 & 4.54 & -0.03 \\
& Wenet testmeeting & 4.73 & 4.70 & -0.03 \\
& FLEURS zh & 2.63 & 2.76 & +0.13 \\
\cmidrule(lr){2-5}
& Average & 3.00 & 3.00 & 0.00 \\
\midrule
\multirow{6}{*}{English} & LibriSpeech clean & 1.40 & 1.38 & -0.02 \\
& LibriSpeech other & 3.14 & 3.16 & +0.02 \\
& Common Voice v11 en & 7.62 & 7.57 & -0.05 \\
& FLEURS en & 3.74 & 3.55 & -0.19 \\
& VoxPopuli cleaned AA & 3.23 & 3.69 & +0.46 \\
\cmidrule(lr){2-5}
& Average & 3.83 & 3.87 & +0.04 \\
\midrule
\multirow{5}{*}{Long-form} & LibriSpeech clean long & 1.27 & 1.27 & 0.00 \\
& LibriSpeech other long & 2.81 & 2.90 & +0.09 \\
& Wenet testnet long & 4.09 & 4.07 & -0.02 \\
& Earnings22 cleaned AA & 6.34 & 6.52 & +0.18 \\
\cmidrule(lr){2-5}
& Average & 3.63 & 3.69 & +0.06 \\
\bottomrule
\end{tabular}
\end{adjustbox}
\end{table}

\section{Conclusion}

We presented \ourmodel, an LLM-based ASR system designed to resolve the fundamental tension between decoder scale and inference latency. By exploiting the inherent determinism of speech transcription, we integrate MTP into the LLM decoder, enabling high-parallelism decoding without compromising recognition accuracy. This design transforms the ASR decoding process from a token-by-token bottleneck into a high-throughput verification regime.
Our technical approach further targets stable long-form transcription and efficient decoding. \ourmodel\ supports a native 32K-context window, allowing the system to process up to 30 minutes of audio in a single pass while maintaining a remarkably low real-time factor. The staged training protocol ensures that the model first establishes a strong autoregressive foundation before specializing in parallel decoding.
Experimental results across Chinese, English, and long-form benchmarks demonstrate that \ourmodel\ establishes a new performance frontier. With an average accepted length of 5.0 tokens per forward step and an RTF of 0.0053 on a single H800 GPU, the system delivers the linguistic depth of a large-scale LLM together with the serving efficiency required for real-time applications. These findings suggest that, for grounded generation tasks like ASR, decoder scaling and acceleration are complementary rather than competing goals when future-token proposals are anchored by the acoustic signal and guarded by autoregressive verification.

\bibliographystyle{waica}
\bibliography{references}

\begin{thebibliography}{10}
\providecommand{\url}[1]{\texttt{#1}}
\providecommand{\urlprefix}{URL }
\providecommand{\doi}[1]{https://doi.org/#1}

\bibitem{an2025funasrtechnicalreport}
An, K., Chen, Y., Chen, Z., Deng, C., Du, Z., Gao, C., et~al.: {Fun-ASR}
  technical report. arXiv preprint arXiv:2509.12508  (2025)

\bibitem{ardila2020commonvoice}
Ardila, R., Branson, M., Davis, K., et~al.: Common voice: A
  massively-multilingual speech corpus. In: Proceedings of the Twelfth Language
  Resources and Evaluation Conference. pp. 4218--4222 (2020)

\bibitem{artificialanalysis2026earnings22cleaned}
{Artificial Analysis}: Earnings22-cleaned-aa: Cleaned ground truth transcripts
  for earnings22 english test set (2026),
  \url{https://artificialanalysis.ai/articles/aa-wer-v2}

\bibitem{artificialanalysis2026voxpopulicleaned}
{Artificial Analysis}: Voxpopuli-cleaned-aa: Cleaned ground truth transcripts
  for voxpopuli english test set (2026),
  \url{https://artificialanalysis.ai/articles/aa-wer-v2}

\bibitem{bai2024seed}
Bai, Y., Chen, J., Chen, J., Chen, W., Chen, Z., Ding, C., Dong, L., Dong, Q.,
  Du, Y., Gao, K., et~al.: Seed-asr: Understanding diverse speech and contexts
  with llm-based speech recognition. arXiv preprint arXiv:2407.04675  (2024)

\bibitem{bain2023whisperx}
Bain, M., Huh, J., Han, T., Zisserman, A.: Whisperx: Time-accurate speech
  transcription of long-form audio. arXiv preprint arXiv:2303.00747  (2023)

\bibitem{bu2017aishell}
Bu, H., Du, J., Na, X., Wu, B., Zheng, H.: {AISHELL-1}: An open-source mandarin
  speech corpus and a speech recognition baseline. In: 2017 20th Conference of
  the Oriental Chapter of the International Coordinating Committee on Speech
  Databases and Speech I/O Systems and Assessment. pp.~1--5 (2017)

\bibitem{cai2024medusa}
Cai, T., Li, Y., Geng, Z., Peng, H., Lee, J.D., Chen, D., Dao, T.: Medusa:
  Simple llm inference acceleration framework with multiple decoding heads.
  arXiv preprint arXiv:2401.10774  (2024)

\bibitem{chan2015listen}
Chan, W., Jaitly, N., Le, Q.V., Vinyals, O.: Listen, attend and spell. arXiv
  preprint arXiv:1508.01211  (2015)

\bibitem{chen2023accelerating}
Chen, C., Borgeaud, S., Irving, G., Lespiau, J.B., Sifre, L., Jumper, J.:
  Accelerating large language model decoding with speculative sampling. arXiv
  preprint arXiv:2302.01318  (2023)

\bibitem{conneau2022fleurs}
Conneau, A., Ma, M., Khanuja, S., et~al.: {FLEURS}: Few-shot learning
  evaluation of universal representations of speech. arXiv preprint
  arXiv:2205.12446  (2022)

\bibitem{deng2025multi}
Deng, Y., Hu, G., Sun, H., Zhang, X., Zhang, H., Tian, F., Yang, X., Yu, G.,
  Chng, E.S.: {Multi-bench}: A multi-turn interactive benchmark for assessing
  emotional intelligence ability of spoken dialogue models. arXiv preprint
  arXiv:2511.00850  (2025)

\bibitem{du2018aishell2}
Du, J., Na, X., Liu, X., Bu, H.: {AISHELL-2}: Transforming {Mandarin} {ASR}
  research into industrial scale. arXiv preprint arXiv:1808.10583  (2018)

\bibitem{fiscus1997rover}
Fiscus, J.G.: A post-processing system to yield reduced word error rates:
  Recognizer output voting error reduction ({ROVER}). In: 1997 IEEE Workshop on
  Automatic Speech Recognition and Understanding Proceedings. pp. 347--354
  (1997)

\bibitem{gloeckle2024betterfaster}
Gloeckle, F., Idrissi, B.Y., Roziere, B., Lopez-Paz, D., Synnaeve, G.: Better
  \& faster large language models via multi-token prediction. arXiv preprint
  arXiv:2404.19737  (2024)

\bibitem{graves2012connectionist}
Graves, A.: Connectionist temporal classification. In: Supervised sequence
  labelling with recurrent neural networks, pp. 61--93. Springer (2012)

\bibitem{graves2012sequence}
Graves, A.: Sequence transduction with recurrent neural networks. arXiv
  preprint arXiv:1211.3711  (2012)

\bibitem{huang2026step35flashopen}
Huang, A., Li, A., Kong, A., Wang, B., Jiao, B., Dong, B., Wang, B., Chen, B.,
  Li, B., Ma, B., et~al.: {Step} 3.5 {Flash}: Open frontier-level intelligence
  with 11{B} active parameters. arXiv preprint arXiv:2602.10604  (2026)

\bibitem{koluguri2024investigating}
Koluguri, N.R., Kriman, S., Zelenfroind, G., Majumdar, S., Rekesh, D., Noroozi,
  V., Balam, J., Ginsburg, B.: Investigating end-to-end asr architectures for
  long form audio transcription. In: ICASSP 2024-2024 IEEE International
  Conference on Acoustics, Speech and Signal Processing (ICASSP). pp.
  13366--13370. IEEE (2024)

\bibitem{leviathan2023fast}
Leviathan, Y., Kalman, M., Matias, Y.: Fast inference from transformers via
  speculative decoding. In: International Conference on Machine Learning. pp.
  19274--19286. PMLR (2023)

\bibitem{lin2026stepaudio}
Lin, B., Zhao, B., Wu, B., Yan, C., Wu, C., Yi, C., Yao, C., Liu, D., Tian, F.,
  Tian, F., et~al.: {StepAudio} 2.5 technical report. arXiv preprint
  arXiv:2605.23463  (2026)

\bibitem{liu2026boosting}
Liu, C., Ma, L., Zhang, X.T., Zhang, Y., Zhang, H., Yang, X., Tian, F.:
  Boosting omni-modal language models: Staged post-training with visually
  debiased evaluation. arXiv preprint arXiv:2605.12034  (2026)

\bibitem{liu2025code}
Liu, H., Zhang, H., Zhang, Q., Zhang, X., Shi, D., Chng, E.S., Li, H.:
  Code-switching speech recognition under the lens: Model-and data-centric
  perspectives. arXiv preprint arXiv:2509.24310  (2025)

\bibitem{malik2021automatic}
Malik, M., Malik, M.K., Mehmood, K., Makhdoom, I.: Automatic speech
  recognition: a survey. Multimedia Tools and Applications  \textbf{80}(6),
  9411--9457 (2021)

\bibitem{panayotov2015librispeech}
Panayotov, V., Chen, G., Povey, D., Khudanpur, S.: {LibriSpeech}: An {ASR}
  corpus based on public domain audio books. In: 2015 IEEE International
  Conference on Acoustics, Speech and Signal Processing. pp. 5206--5210 (2015)

\bibitem{park2019specaugment}
Park, D.S., Chan, W., Zhang, Y., et~al.: {SpecAugment}: A simple data
  augmentation method for automatic speech recognition. In: Interspeech 2019.
  pp. 2613--2617 (2019)

\bibitem{peng2026vibevoiceasrtechnicalreport}
Peng, Z., Yu, J., Chang, Y., Wang, Z., Dong, L., Hao, Y., et~al.:
  {VIBEVOICE-ASR} technical report. arXiv preprint arXiv:2601.18184  (2026)

\bibitem{radford2023robust}
Radford, A., Kim, J.W., Xu, T., Brockman, G., McLeavey, C., Sutskever, I.:
  Robust speech recognition via large-scale weak supervision. In: International
  conference on machine learning. pp. 28492--28518. PMLR (2023)

\bibitem{shi2026qwen3asrtechnicalreport}
Shi, X., Wang, X., Guo, Z., Wang, Y., Zhang, P., Zhang, X., et~al.: {Qwen3-ASR}
  technical report. arXiv preprint arXiv:2601.21337  (2026)

\bibitem{tian2025step}
Tian, F., Zhang, X.T., Zhang, Y., Zhang, H., Li, Y., Liu, D., Deng, Y., Wu, D.,
  Chen, J., Zhao, L., et~al.: {Step-Audio-R1} technical report. arXiv preprint
  arXiv:2511.15848  (2025)

\bibitem{wu2025stepaudio2technicalreport}
Wu, B., Yan, C., Hu, C., Yi, C., Feng, C., Tian, F., Shen, F., Yu, G., Zhang,
  H., Li, J., et~al.: {StepAudio} 2 technical report. arXiv preprint
  arXiv:2507.16632  (2025)

\bibitem{wu2025mind}
Wu, D., Zhang, H., Chen, J., Liu, H., Chng, E.S., Tian, F., Yang, X., Zhang,
  X., Jiang, D., Yu, G., et~al.: Mind-paced speaking: A dual-brain approach to
  real-time reasoning in spoken language models. arXiv preprint
  arXiv:2510.09592  (2025)

\bibitem{xu2025qwen3omnitechnicalreport}
Xu, J., Guo, Z., Hu, H., Chu, Y., Wang, X., He, J., Wang, Y., Shi, X., He, T.,
  Zhu, X., et~al.: {Qwen3-Omni} technical report. arXiv preprint
  arXiv:2509.17765  (2025)

\bibitem{xuan2023new}
Xuan, Y., Zhang, X., Li, S.S., Shen, Z., Xie, X., Garcia, L.P., Togneri, R.: A
  new approach to extract fetal electrocardiogram using affine combination of
  adaptive filters. In: ICASSP 2023-2023 IEEE International Conference on
  Acoustics, Speech and Signal Processing (ICASSP). pp.~1--5. IEEE (2023)

\bibitem{zeng2023improving}
Zeng, C., Wang, X., Miao, X., Cooper, E., Yamagishi, J.: Improving
  generalization ability of countermeasures for new mismatch scenario by
  combining multiple advanced regularization terms. In: Proc. Interspeech 2023.
  pp. 1998--2002 (2023)

\bibitem{zhang2022wenetspeech}
Zhang, B., Lv, H., Guo, P., Shao, Q., Yang, C., Xie, L., Xu, X., Bu, H., Chen,
  X., Zeng, C., et~al.: {WenetSpeech}: A 10000+ hours multi-domain mandarin
  corpus for speech recognition. In: ICASSP 2022 - 2022 IEEE International
  Conference on Acoustics, Speech and Signal Processing. pp. 6182--6186 (2022)

\bibitem{zhang2026duplexsla}
Zhang, H., Chen, J., Wu, D., Li, Y., Zhang, Y., Zhang, X.T., Liu, C., Lin, Q.,
  Peng, Y., Liu, H., et~al.: {DuplexSLA}: A full-duplex spoken language model
  with synchronized speech, language, and action. arXiv preprint
  arXiv:2605.20755  (2026)

\bibitem{zhang2025impact}
Zhang, H., Liu, H., Zhang, X., Zhang, Q., Hu, Y., Zhao, J., Tian, F., Yang, X.,
  Garcia, L.P., Chng, E.S.: Impact of frame rates on speech tokenizer: A case
  study on mandarin and english. arXiv preprint arXiv:2505.17076  (2025)

\bibitem{zhang2026trap}
Zhang, X., Li, Y., Zhang, H., Han, S., Liu, H., Zhang, Q., Ahmed, B., Epps, J.:
  The {WER} trap: Shattering the illusion of unified tokens in speech language
  models. arXiv preprint arXiv:2605.29209  (2026)

\bibitem{zhang2025mamba}
Zhang, X., Zhang, Q., Liu, H., Xiao, T., Qian, X., Ahmed, B., Ambikairajah, E.,
  Li, H., Epps, J.: Mamba in speech: Towards an alternative to self-attention.
  IEEE Transactions on Audio, Speech and Language Processing  (2025)

\bibitem{zhang2026step}
Zhang, Y., Zhang, X.T., Liu, D., Tian, F., Deng, Y., Chen, J., Lin, Q., Zhang,
  H., Li, Y., Gong, J., et~al.: {Step-Audio-R1.5} technical report. arXiv
  preprint arXiv:2604.25719  (2026)

\end{thebibliography}

\end{document}